\documentclass[12pt, preprint]{aastex}
\usepackage{lineno}

\usepackage{natbib}

\shorttitle{Fermi Observations of  Cas A}
\slugcomment{Submitted to ApJL. V6.0 draft (\today )}

\begin{document}

\title{{\emph{Fermi}}-LAT discovery of GeV gamma-ray emission from the
  young supernova remnant Cassiopeia A}

\author{
A.~A.~Abdo\altaffilmark{2,3}, 
M.~Ackermann\altaffilmark{4}, 
M.~Ajello\altaffilmark{4}, 
A.~Allafort\altaffilmark{4}, 
L.~Baldini\altaffilmark{5}, 
J.~Ballet\altaffilmark{6}, 
G.~Barbiellini\altaffilmark{7,8}, 
M.~G.~Baring\altaffilmark{9}, 
D.~Bastieri\altaffilmark{10,11}, 
B.~M.~Baughman\altaffilmark{12}
% ...
}
% ...
\altaffiltext{2}{Space Science Division, Naval Research Laboratory, Washington, DC 20375, USA}
\altaffiltext{3}{National Research Council Research Associate, National Academy of Sciences, Washington, DC 20001, USA}
\altaffiltext{4}{W. W. Hansen Experimental Physics Laboratory, Kavli Institute for Particle Astrophysics and Cosmology, Department of Physics and SLAC National Accelerator Laboratory, Stanford University, Stanford, CA 94305, USA}
\altaffiltext{5}{Istituto Nazionale di Fisica Nucleare, Sezione di Pisa, I-56127 Pisa, Italy}
\altaffiltext{6}{Laboratoire AIM, CEA-IRFU/CNRS/Universit\'e Paris Diderot, Service d'Astrophysique, CEA Saclay, 91191 Gif sur Yvette, France}
\altaffiltext{7}{Istituto Nazionale di Fisica Nucleare, Sezione di Trieste, I-34127 Trieste, Italy}
\altaffiltext{8}{Dipartimento di Fisica, Universit\`a di Trieste, I-34127 Trieste, Italy}
\altaffiltext{9}{Rice University, Department of Physics and Astronomy, MS-108, P. O. Box 1892, Houston, TX 77251, USA}
\altaffiltext{10}{Istituto Nazionale di Fisica Nucleare, Sezione di Padova, I-35131 Padova, Italy}
\altaffiltext{11}{Dipartimento di Fisica ``G. Galilei", Universit\`a di Padova, I-35131 Padova, Italy}
\altaffiltext{12}{Department of Physics, Center for Cosmology and Astro-Particle Physics, The Ohio State University, Columbus, OH 43210, USA}
% ...

\begin{abstract}
  We report on the first detection of GeV high-energy gamma-ray
  emission from a young supernova remnant with the Large Area
  Telescope aboard the {\emph{Fermi Gamma-ray Space Telescope}}. These
  observations reveal a source with no discernible spatial extension
  detected at a significance level of 12.2$\sigma$  above
  500~MeV at a location that is consistent with the position of the
  remnant of the supernova explosion that occurred around 1680 in the
  Cassiopeia constellation -- Cassiopeia~A. The gamma-ray flux and
  spectral shape of the source are consistent with a scenario in which
  the gamma-ray emission originates from relativistic particles
  accelerated in the shell of this remnant.  The total content of
  cosmic rays (electrons and protons) 
  accelerated in Cas~A can be estimated as $W_{\rm CR}
  \simeq (1\mbox{--}4) \times 10^{49}\ {\rm erg}$
  thanks to the well-known density in the remnant  
  assuming that the observed gamma-ray originates in the SNR shell(s).
  The magnetic field in the radio-emitting plasma can be robustly
  constrained as $B \geq 0.1\ {\rm mG}$, providing new evidence of the
  magnetic field amplification at the forward shock and the strong
  field in the shocked ejecta.
\end{abstract}


\section{Introduction}
Supernova remnants (SNRs) have long been considered as the primary
candidates for the origin of Galactic Cosmic Rays (CRs).
Specifically, diffusive shock
acceleration~\citep{Bell-1,BlandfordOstriker,Jones,MalkovDrury} is
widely accepted as the mechanism by which charged particles can be
accelerated to very high energies at collisionless shocks driven by
supernova explosions.  To maintain the energy density of the Galactic
CRs, the kinetic energy released in supernova explosions has to be
efficiently transferred to CRs with a conversion efficiency of $\sim
10\%$~\citep{Ginzburg}.  However, it has not yet been confirmed
whether strong shock waves in SNRs are indeed capable of transferring
this amount of energy into the acceleration of CR ions, and in doing
so generating a CR energy density comparable to the energy density
contained in the expansion ram pressure of the supernova.

Shock energy can be converted to another form of energy: turbulent
magnetic fields.  Recent theoretical work indicates that the
magnetic field can be largely amplified by streaming of CRs
themselves, as an integral part of efficient CR acceleration at
collisionless shocks \citep{BellLucek, Bell}.  Observations of
synchrotron X-ray emission in young SNRs have shown that the magnetic
field at supernova shocks can be amplified way beyond the factor of
$\sim 4$ that is expected for standard compression of interstellar
magnetic fields in the absence of CR modification of the shock's
hydrodynamic
structure~\citep{Uchiyama2007,Voelk2005,Bamba2005,Vink2003}.  The
amplified magnetic field would allow acceleration of protons up to the
\emph{knee} in the CR spectrum at $\sim 10^{15}$~eV, the presumed
endpoint of the Galactic component \citep{BellLucek}.

Cassiopeia~A (Cas~A) is the remnant of one of a handful of historical
supernovae; the explosion around AD 1680 that gave rise to Cas~A has
probably been the last Galactic supernova witnessed by
humans~\citep{Ashworth1980}. It is the brightest radio source in our
Galaxy \citep{Baars} and its overall brightness across the
electromagnetic spectrum makes it a unique laboratory for studying
high-energy phenomena in SNRs.  Cas~A has an angular size of
$2.5\arcmin$ in radius corresponding to a physical size of 2.34~pc at
a distance of 3.4$^{+0.3}_{-0.1}$~kpc~\citep{Reed1995}.
Cas~A was the first SNR detected in TeV gamma rays, first by
HEGRA~\citep{HEGRA:casA} and later confirmed by
MAGIC~\citep{MAGIC:casA} and VERITAS~\citep{VERITAS:casA}, firmly
establishing the existence of multi-TeV particles.  These
higher-energy gamma rays are generally attributed to particles
accelerated in the shock waves of the SNR, even though the TeV
gamma-ray source is not resolved. The emission mechanism(s)
  responsible for the TeV gamma-ray emission remained unsettled. 
EGRET reported only upper limits on the gamma-ray emission in the GeV
band~\citep{Esposito1996}.

The advent of the Large Area Telescope (LAT) on board the \emph{Fermi}
Gamma-ray Space Telescope for the first time makes it possible to
detect GeV gamma rays from the shell of Cas~A.  The relativistic
bremsstrahlung of accelerated electrons has been predicted to lie
above the sensitivity of the LAT \citep{Atoyan2000}.  The prediction
of the bremsstrahlung flux is controlled by the strength of the
magnetic field and therefore the gamma-ray flux provides a measure of
the amplified magnetic field in the supernova
shell~\citep{Cowsik}. Inverse Compton (IC) scattering is another
important mechanism of gamma-ray production by high-energy electrons;
seed photons that can be upscattered to gamma-rays include the
interstellar radiation field, the cosmic microwave background (CMB),
and the far-infrared (FIR) radiation by Cas~A itself. In addition to
these leptonic emissions, the decay of $\pi^0$-mesons produced in
inelastic collisions between high-energy protons (and nuclei) and
background gas may contribute to the GeV gamma-ray flux. The
importance of the GeV observations of young SNRs is emphasized by the
fact that the energy density of the accelerated particles
is measurable thanks to the well-constrained gas and radiation density
in Cas A in addition to that of the magnetic field.  In this paper, we
report the discovery of GeV gamma-ray emission coincident with Cas~A
based on the first year observations with the \emph{Fermi} LAT.

\section{Observations}

The LAT onboard \emph{Fermi} is a pair-conversion gamma-ray detector
operating between 20~MeV and 300~GeV. 
The LAT has a wide field of view of $\sim 2.4$~sr at 1 GeV, and
observes the entire sky every 2 orbits ($\sim 3$ hr for \emph{Fermi}'s
orbit at an altitude of $\sim 565$ km).  The full details of the
instruments are given in \citet{LATPaper}.

A total exposure of $3\times 10^{10}\ {\rm cm}^2\ {\rm s}$ (at 1 GeV)
has been obtained for Cas~A during the period between August 4th 2008
and September 4th 2009, corresponding to 394 days of observations.
The data analysis was performed using the LAT Science Tools package
with the {\emph{P6\_V3}} post-launch instrument response
function~\citep{Rando2009}. The standard event selection for source
analysis, resulting in the strongest background-rejection power
({\emph{diffuse}} event class) was applied. In addition photons coming
from zenith angles larger than $105^{\circ}$ were rejected to reduce
the background from gamma rays produced in the atmosphere of the
Earth. The analysis was further restricted to the energy range above
200~MeV where the uncertainties in the effective area become smaller
than 10\%.

\section{Analysis and Results}

The analysis of the gamma-ray event data was performed using {\it
  gtlike}, which is available as part of \emph{Fermi} Science
Tools\footnote{Software and documentation of the \emph{Fermi} Science
  Tools are distributed by the Fermi Science Support Center at
  http://fermi.gsfc.nasa.gov/ssc}.  The tool {\it gtlike} employs a
maximum likelihood technique to assess the statistical significance of
sources and to estimate spectral parameters~\citep{Mattox}.  The
background gamma-ray model includes background sources from the
11-month catalog of Fermi-LAT sources at fixed coordinates, the
galactic diffuse emission (using gll\_iem\_v02.fit) and an isotropic
component (using isotropic\_iem\_v02.txt).  The maximum likelihood
analysis is performed inside a region-of-interest (ROI) of $10\degr$
radius centered on Cas~A.  The independent tool {\it sourcelike} was
employed to determine the source position and constrain the angular
extent of the source.  {\it Sourcelike} performs a maximum likelihood
fit to simultaneously optimize position and extent of the source
(given an assumption about the shape of the source) in independent
energy bands.
{\it Sourcelike} can also be used to assess the test-statistic (TS)
value and to compute the spectra of both extended and point-sources.
The position determined with {\it sourcelike} is consistent with what
was found by the automated LAT catalog tool.  The spectral parameters
between {\it sourcelike} and {\it gtlike} for this best-fit position
are found to be consistent, yielding an independent check of the main
results of the paper.

\begin{figure}[h]
  \centering
  \includegraphics[width=0.8\textwidth]{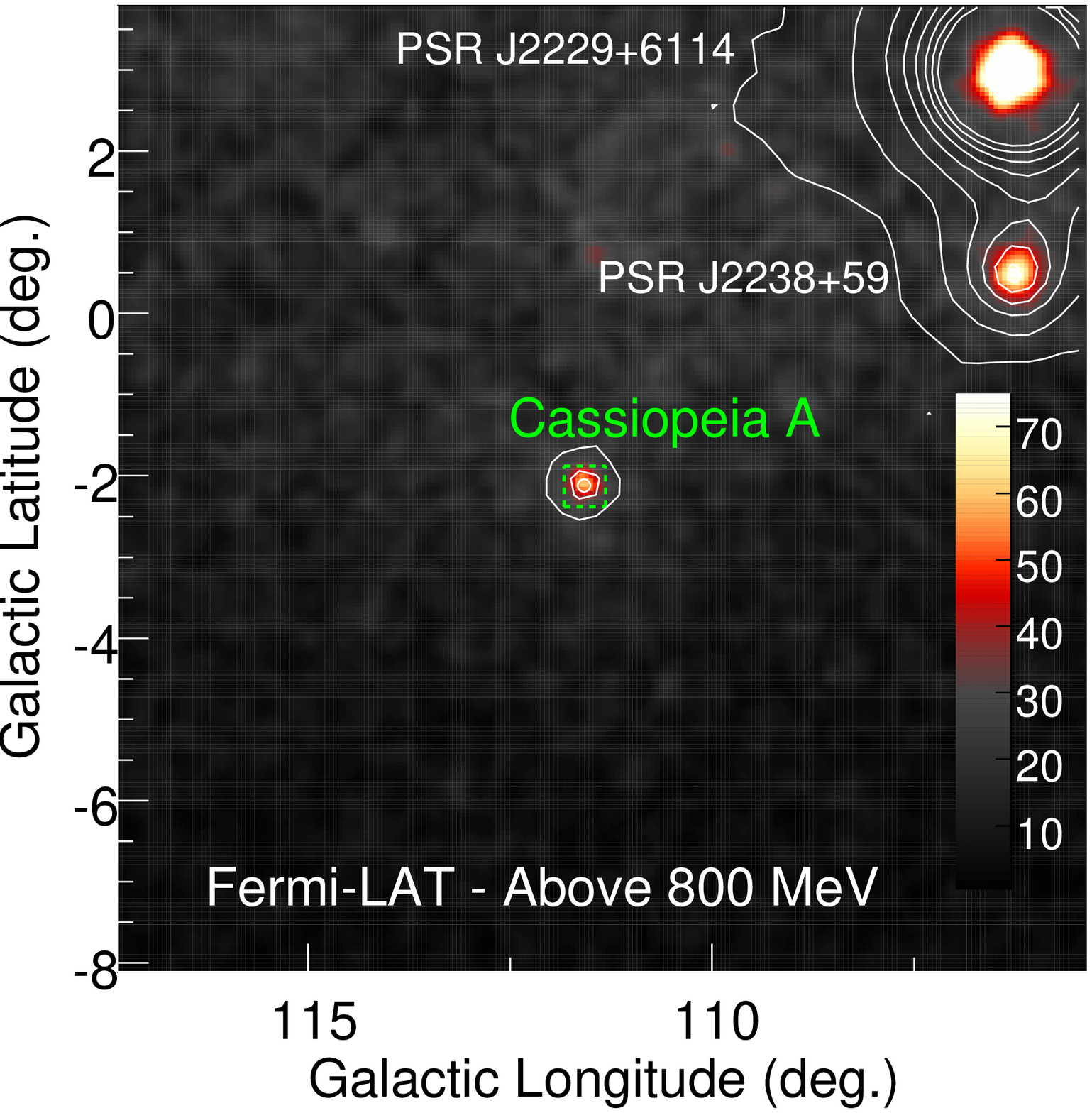}
  \caption{Smoothed gamma-ray count map of the region surrounding
    Cas~A with side-length 12$^{\circ}$, binned in square pixels of
    side-length 0.05$^{\circ}$. Smoothing was done with a Gaussian of
    width 0.1$^{\circ}$. Overlaid are test statistics contours (${\rm
      TS} = -2 \ln(-L)$) for the assumption of a point-source located
    at the position of the pixel. Contours of 25, 100, 200, 300, 400,
    500, 600 and 700 TS levels are shown. Only events with energies
    larger than 800~MeV were used.}
  \label{fig:Map}
\end{figure}

\begin{figure}[h]
  \centering
  \includegraphics[width=0.8\textwidth]{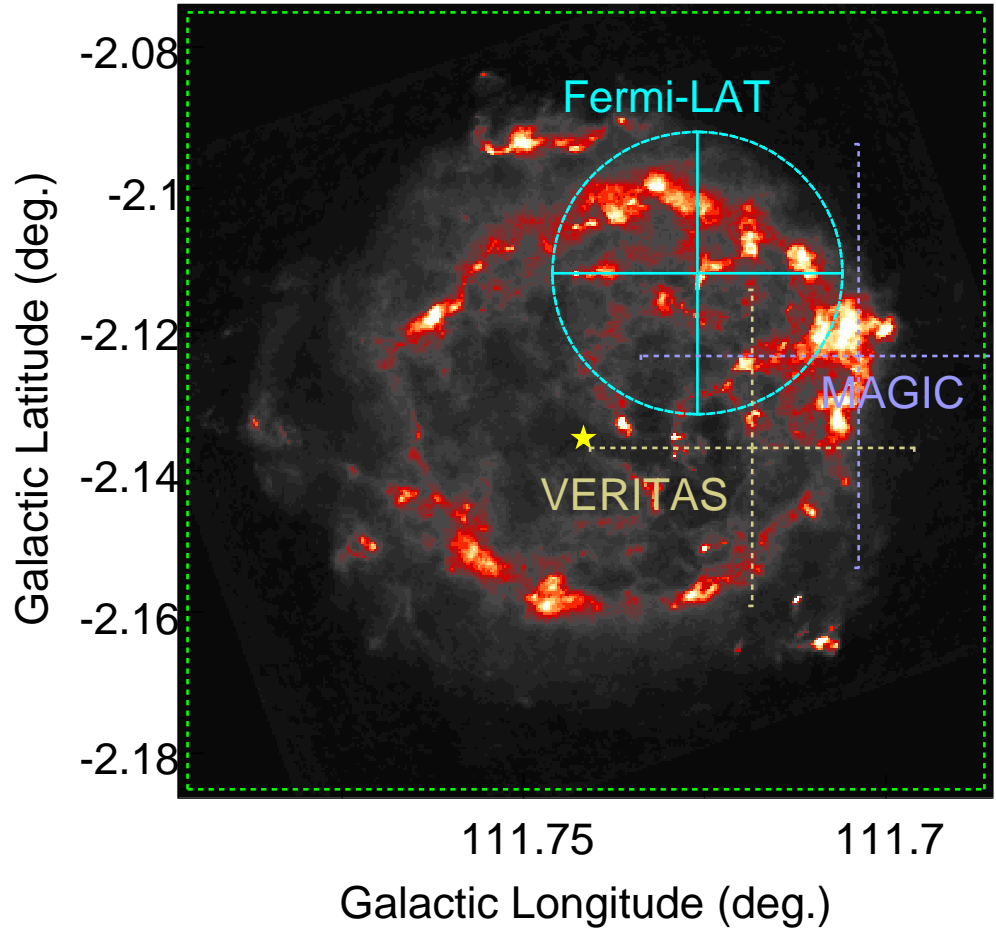}
  \caption{The VLA 20~cm radio map of the Cas~A supernova
    shell~\citep{AndersonRudnick}. Shown is the region corresponding
    to the green square in Figure~\ref{fig:Map}. The graph shows the
    GeV position with error bars (conservatively adding statistical
    and systematic errors in quadrature) as a cyan circle and the
    position of the CCO as a yellow star. Also shown in the plot are
    the best-fit positions for MAGIC~\citep{MAGIC:casA} and
    VERITAS~\citep{VERITAS:casA}.}
  \label{fig:Positions}
\end{figure}

The analysis clearly shows a source above the background coincident
with the SNR Cas~A (see Fig.~\ref{fig:Map}).  The source is detected
at a significance level of 12.2 $\sigma$ (or a TS value of 148) above
the background at a best fit position of $\alpha_{2000} =
23\mathrm{h}23\mathrm{m}17.5\mathrm{s}, \delta_{2000} =
58\mathrm{h}49\mathrm{m}43.1\mathrm{s}$ with a statistical uncertainty
on the location of $1.0\arcmin$ (68\% confidence level).  The
systematic error in the position due to alignment of the telescope
system and inaccurate description of the point-spread function of the
instrument is estimated to be $0.3\arcmin$. The position is spatially
coincident with Cas~A and with the MAGIC and VERITAS-detected TeV
gamma-ray source~\citep{MAGIC:casA,VERITAS:casA} as shown in
Fig.~\ref{fig:Positions}.  An upper limit on the size of the gamma-ray
emission has been obtained in a maximum likelihood fit by
investigating the decrease of the likelihood with increasing source
size. Under the assumption of a Gaussian shape this 68\% upper limit
amounts to $3.5\arcmin$ (1-$\sigma$ radius).  This value is larger
than the radius of the SNR ($2.5\arcmin$), which thus currently
cannot be resolved by the {\emph{Fermi}}-LAT. A check for variability
in this data set with one-month time bins shows no sign for any change
in the gamma-ray flux with time, indicating a steady source of
emission.

A spectral analysis of the gamma-ray emission has been performed in an
energy range of 200 MeV--50 GeV.  The source is detected only above
500~MeV. The spectrum shown in Fig.~\ref{fig:leptons} reveals a
relatively flat energy distribution and can be fitted by a power law
with spectral index of $\Gamma = 2.0 \pm 0.1$ between 0.5 and 50 GeV.
The systematic uncertainty in the spectral index determination from
the uncertainty in the normalisation of the Galactic diffuse emission
(conservatively estimated to be 10\%) amounts to $\sim 0.1$. The
integral photon flux above 500~MeV amounts to $(8.7\pm 1.3)\times
10^{-9}\ {\rm photons}\ {\rm cm}^{-2}\ {\rm s}^{-1}$.  A
likelihood-ratio test was performed to check the presence of a
spectral cutoff using a spectral function of $dN/dE = K E^{-\Gamma}
\exp (-E/E_{\rm cut} )$.  The presence of the exponential cutoff is
not statistically significant, given the resultant likelihood ratio of
$-2\ln (L_{\rm PL}/L_{\rm cutoff}) = 2.6$.  We formally report the
best-fit parameters of the cutoff power-law model to make a comparison
with LAT pulsars: $\Gamma = 1.7\pm 0.2$ and $E_{\rm cut} = 16\pm 9\
{\rm GeV}$.

\section{Discussion}

The detection of gamma-ray emission from the direction of Cas~A raises
the following questions: ``Is there a compact source such as a pulsar
dominating the emission?"  and ``What is the radiating particle
population responsible for the emission?".  Two emission scenarios
seem plausible: emission from the central compact object (CCO) in
Cas~A or emission generated in the SNR. Since the GeV gamma-ray source
coinciding with Cas~A is a point-source in the LAT, these two
scenarios cannot be distinguished on positional grounds therefore
circumstantial evidence has to be considered.

The point-like central X-ray source \citep{Tananbaum1999} is generally
thought to be the left-over of the explosion of the massive progenitor
star \citep{Pavlov2000} and known as a CCO given that it is radio
quiet, un-pulsed in X-rays, and has an X-ray spectrum described by
blackbody with characteristic temperatures of about 0.4 keV without
indication of a non-thermal component.  A recent {\emph{Chandra}}
observation resulted in a 3$\sigma$ limit on the pulsed fraction of
16\% for periods larger than 0.68~s~\citep{Pavlov2009}.  The fact that
no pulsation has been detected at any waveband for the CCO does not
rule out, that the neutron star is emitting gamma rays. The
\emph{Fermi}-LAT has detected several neutron stars as gamma-ray-only
pulsars, pulsars not previously known from observations in other
wavebands~\citep{FermiCTA1,FermiBlindPulsars}.

We searched for gamma-ray pulsations from the source coincident with
Cas~A using the full data set.  We applied the standard
time-differencing technique~\citep{BlindTechnique}, using a maximum
frequency of 64 Hz, and a long time-difference window of $\sim$12
days. We found no evidence for pulsations which makes the neutron star
scenario less likely. 
A comparison of the blind search pulsars discovered so
far~\citep{FermiBlindPulsars} and the known radio pulsars detected by
the LAT, suggests that the blind search is approximately 2--3 times
less sensitive than a standard pulsation search using the known timing
solution. This results in a 5$\sigma$ limit on the pulsed flux of
$\sim2\times10^{-7}\ {\rm photons}\ {\rm cm}^{-2}\ {\rm s}^{-1}$
\citep{FermiPulsarCat}. In addition, there is no GeV gamma ray source
in the 1st Fermi-LAT source catalog that is associated with a known
CCO.

Furthermore, the spectrum does not support a pulsar hypothesis. The
energy spectra of pulsars are usually flat at energies below 1~GeV and
show exponential cutoffs in the energy range between 1~GeV and
8~GeV~\citep{FermiPulsarCat}.  These characteristics do not mirror the
LAT spectrum which is best described by a power-law shape with an
index of $\Gamma = 2.0\pm 0.1$ up to 50~GeV with no significant sign
of a high-energy cutoff.  A formal fit with an exponential cutoff
model yields $E_{\rm cut} = 16\pm 9\ {\rm GeV}$.  This is rather high
for a gamma-ray pulsar; no LAT pulsars show $E_{\rm cut} > 8\ {\rm
  GeV}$ so far \citep{FermiBlindPulsars}.

The scenario in which the GeV gamma rays are emitted in the shell of
the SNR is therefore favored.  The gamma-ray emission could be
produced by electrons accelerated at the forward shock through
relativistic bremsstrahlung or IC.  Alternatively, the GeV gamma-ray
emission could be predominantly produced by accelerated hadrons
through interaction with the background gas and subsequent
$\pi^0$-decay.  Recent studies showed electron acceleration to
multi-TeV energies is likely to take place also at the reverse shock
in the supernova ejecta \citep{Uchiyama2008,HelderVink}.

\begin{figure}[h]
  \centering
 \includegraphics[width=0.8\textwidth]{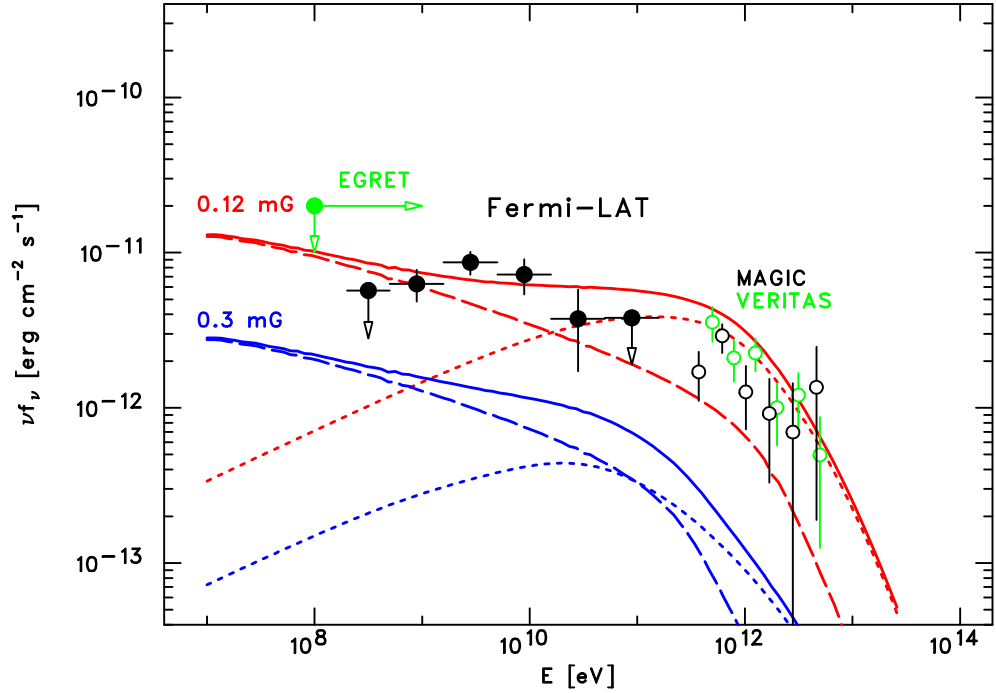}
 \caption{Energy spectrum of Cas~A in a leptonic emission model. Shown
   is the {\emph{Fermi}}-LAT detected emission (filled circles) in
   comparison to the energy spectra detected by MAGIC \citep[black
   open circles;][]{MAGIC:casA} and VERITAS \citep[green open
   circles;][]{VERITAS:casA}.  The red curves show a leptonic model
   calculated for $B=0.12\ {\rm mG}$ while the blue curves show one
   for $B=0.3\ {\rm mG}$.  Dashed lines show the bremsstrahlung
   components with $n_{\rm eff} \simeq 26\ {\rm cm^{-3}}$, and dotted
   lines show the IC component.}
 \label{fig:leptons}
\end{figure}

First, we consider the shocked circumstellar region between the
contact discontinuity and the forward shock
\citep[see][]{Gotthelf2001}, assumed to have constant magnetic field
of $B$ (a free parameter) and shocked circumstellar medium with a
constant density of $n_{\rm H} = 10\ {\rm cm^{-3}}$
  \citep{LamingHwang2003}. Electrons are accelerated to multi-TeV
energies at the forward shock as traced by synchrotron X-ray outer
filaments \citep{Hughes}.  We adopt here an electron acceleration
spectrum $Q_{\rm e}(E) \propto E^{-2.34} \exp( -E/E_{\rm m} )$ to
match the radio-IR spectral index of $\alpha = 0.67$ \citep{Rho},
since both the GeV gamma-ray emission and the radio synchrotron
emission sample similar electron energies.  About half of the total
radio flux is attributable to this region (so-called {\it plateau}),
while another half is to the reverse shock region (so-called {\it
  bright ring}; see below). Given the radio flux and the effective
density ($n_{\rm eff} = \Sigma n_i Z_i(Z_i+1) \simeq 26\ {\rm
  cm^{-3}}$), the flux of bremsstrahlung is controlled by $B$.  In
Figure~\ref{fig:leptons}, we show a leptonic model with $B=0.12\ {\rm
  mG}$ (red curves), which can broadly explain the observed GeV flux.
The maximum energy is set to be $E_{\rm m} = 40\ {\rm TeV}$
\citep{Vink2003}.
Shown are contributions from bremsstrahlung (dashed) and from IC
scattering (dotted) produced by accelerated electrons that suffer
synchrotron cooling at high energies.  The bremsstrahlung spectrum
consisting of electron-ion and electron-electron components is
computed as in~\citet{Baring1999}.  The radiation field for the IC
component is dominated by FIR emission from the Cas~A ejecta,
characterized by a temperature of 100~K and an energy density of $\sim
2\ \rm{eV}\ \rm{cm}^{-3}$\citep{Mezger1986}, a factor of 8 larger
  than the energy density in the CMB.  The IC/FIR emission exceeds
  IC/CMB by a factor of 2.7 at a gamma-ray energy of 10 GeV.  The
  value of $B=0.12\ {\rm mG}$ is consistent with $B=0.08\mbox{--}0.16\
  {\rm mG}$ at the forward shock estimated by \citet{Vink2003} based on the width of a
  synchrotron X-ray filament.  Note however that a somewhat higher
  value of $B \simeq 0.3\ {\rm mG}$ was obtained by
  \citet{Parizot2006} using the same filament width but including
  projection effects. The total amount of electrons in this case is
$W_e (>10\ {\rm MeV}) \simeq 1\times 10^{49}\ {\rm erg}$.  Also shown
in Fig.~\ref{fig:leptons} is the case of $B=0.3\ {\rm mG}$ (blue
curves), which predicts a lower gamma-ray flux than the observed one.

\begin{figure}[h]
  \centering
 \includegraphics[width=0.8\textwidth]{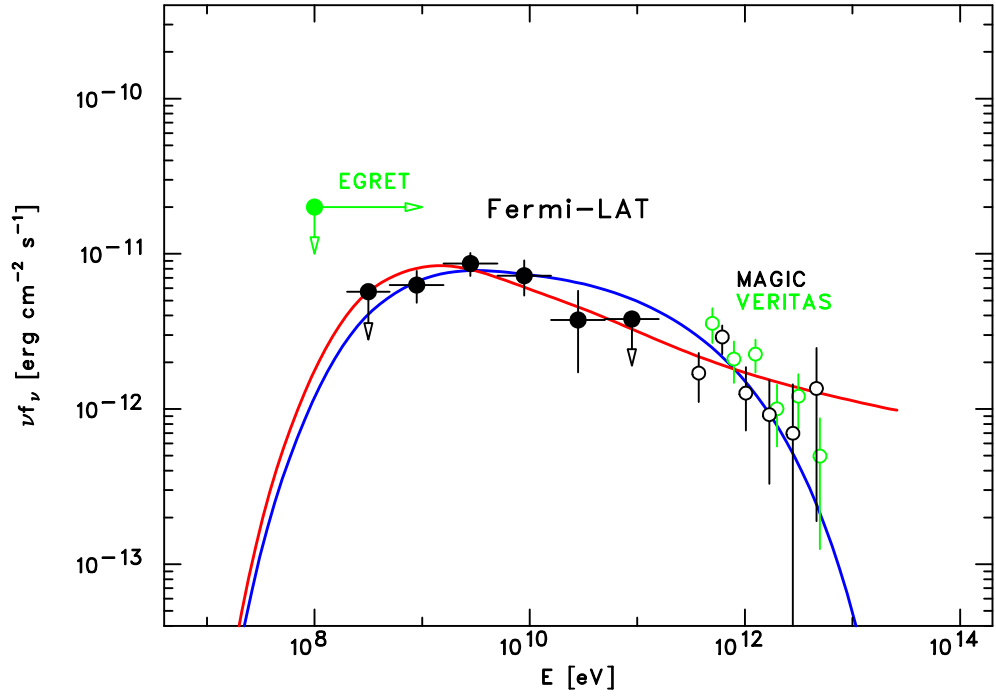}
 \caption{Same as Fig.~\ref{fig:leptons} but in a hadronic emission
   model.  Shown are $\pi^0$-decay spectra for two possible proton
   spectra.  The blue line shows a harder proton index of 2.1 with a
   cutoff energy of 10~TeV ($W_p = 3.2 \times 10^{49}\ \rm{erg}$ for
   $n_{\rm H} = 10\ {\rm cm}^{-3}$).  The red line shows a softer
   proton index of 2.3 without a cutoff.}
  \label{fig:hadrons}
\end{figure}

In a scenario in which the gamma-rays are generated by $\pi^0$-decay
of accelerated hadrons, the gamma-ray spectrum can be well matched
with either the proton acceleration spectrum $Q_{\rm p}(p) \propto
p^{-2.3}$ (a red curve in Fig.~\ref{fig:hadrons}), or a harder proton
spectrum of $Q_{\rm p}(p)\propto p^{-2.1}$ with an exponential cutoff
at 10~TeV that is arbitrary introduced (blue curve).  Here $p$
denotes momentum of accelerated protons.  The gamma-ray spectrum is
calculated following \citet{Kamae2006} with a scaling factor of 1.85
for helium and heavy nuclei \citep{Mori2009}.  The total proton
content amounts to $W_p (>10\ {\rm MeV}\,c^{-1}) \simeq 3.8\times
10^{49}\ {\rm erg}$ in the case of the softer spectrum and to $W_p
(>10\ {\rm MeV}\,c^{-1}) \simeq 3.2\times 10^{49}\ {\rm erg}$ in the
case of the harder proton spectrum with the cutoff.  In both cases the
energy content corresponds to less than 2\% of the estimated explosion
kinetic energy of $E_{\rm sn} = 2\times 10^{51}\ {\rm erg}$.
Therefore, the cosmic-ray pressure would not be large enough to change
the hydrodynamics of Cas~A. This is consistent with
  \emph{Chandra} X-ray measurements of the remnant's spatial structure
  \citep[e.g.,][]{Gotthelf2001}; the ratio of the radii of forward and
  reverse shocks can be reproduced by hydrodynamical models that do
  not include cosmic-ray acceleration \citep{LamingHwang2003}.

Comparing the leptonic and hadronic models, it seems clear that the
hadronic scenario can better fit the data due to the turnover at low
energies that is not well reproduced in the leptonic scenario. Given
the uncertainties in the diffuse model that have stronger effects at
the low-energy end, we refrain, however, from strong claims about the
radiating particle population at this point. Changing the diffuse
model normalisation by $\pm 10\%$ (a conservative assumption on the
uncertainty) largely affects the energy points at and below 1~GeV. The
resulting effect on the flux point at 1~GeV is a 25\% upward and 65\%
downward shift. A more detailed investigation of the lower energy end
of the LAT spectrum will be possible with future LAT data.

The shocked ejecta gas that emits strong radio and infrared
synchrotron light (known as the {\it bright ring}) is another
potential gamma-ray emitting region.  Using $M_{\rm ejecta} =
2M_{\odot}$ \citep{Willingale, LamingHwang2003} comprised of only
oxygen, we obtain $n_{\rm eff} = n_{\rm O} Z_{\rm O}(Z_{\rm O}+1)
\simeq 32\ {\rm cm^{-3}}$.  This happens to be similar to the value of
$n_{\rm eff}$ in the forward shock region.  Also, the baryon density
of $n_{\rm O}A_{\rm O} \simeq 7\ {\rm cm^{-3}}$ is close to that in
the forward shock region.  Therefore, the $\pi^0$-decay predictions
are essentially the same as those described above, though the energy
budget is tighter in the reverse shock. The total thermal energy
stored in the shocked ejecta would be only $\sim 1\times 10^{50}\ {\rm
  erg}$.  On the other hand, a combination of $n_{\rm eff} \simeq 32\
{\rm cm^{-3}}$ and the large magnetic field in the shocked ejecta ($B
\ga 0.5 \ {\rm mG}$) inferred by X-ray variability
\citep{Uchiyama2008} makes it difficult to attribute the gamma-ray
emission to the relativistic bremsstrahlung in the shocked ejecta.

Regardless of the origin(s) of the observed gamma rays, the total
content of CRs accelerated in Cas~A can be obtained as $W_{\rm CR} =
W_e+W_p \simeq (1\mbox{--}4) \times 10^{49}\ {\rm erg}$, and the
magnetic field amplified at the shock and the field in the shocked
ejecta can be constrained as $B \geq 0.12\ {\rm mG}$.  Even though
Cas~A is considered to have entered the Sedov phase, the total amount
of CRs accelerated in the remnant constitutes only a minor fraction
($\la 2\%$) of the total kinetic energy of the supernova.  The
bremsstrahlung spectrum and $\pi^0$-decay spectrum have rather
different predictions below 1 GeV.  The hard spectrum below 1 GeV
would favor the $\pi^0$-decay origin, though the current LAT data
quality does not rule out the bremsstrahlung model.

\acknowledgments
The {\it Fermi} LAT Collaboration acknowledges support from a number
of agencies and institutes for both development and the operation of
the LAT as well as scientific data analysis. These include NASA and
DOE in the United States, CEA/Irfu and IN2P3/CNRS in France, ASI and
INFN in Italy, MEXT, KEK, and JAXA in Japan, and the K.~A.~Wallenberg
Foundation, the Swedish Research Council and the National Space Board
in Sweden. Additional support from INAF in Italy and CNES in France
for science analysis during the operations phase is also gratefully
acknowledged.


\end{document}